\documentclass[aps,preprintnumbers,nofootinbib,floatfix]{revtex4}
\usepackage{amssymb}
\usepackage{bm}
\usepackage{graphicx}

\begin{document}

\title{IMAGES OF QUARK INTRINSIC MOTION IN COVARIANT PARTON MODEL}
\author{A.V.~Efremov$^{1}$, P. Schweitzer$^{2}$, O.~V.~Teryaev$^{1}$, P.
Zavada$^{3}$}
\affiliation{{\ }$^{1}$ {Bogoliubov Laboratory of Theoretical Physics, JINR, 141980
Dubna, Russia}\\
$^{2}$ {Department of Physics, University of Connecticut, Storrs, CT 06269,
U.S.A.}\\
$^{3}$ {Institute of Physics AS CR, Na Slovance 2, CZ-182 21 Prague 8, Czech
Rep.}}

\begin{abstract}
\noindent We discuss the relations between TMDs and PDFs in the framework of
the covariant parton model. The quark OAM and its connection to TMDs are
studied as well.
\end{abstract}

\maketitle

\section{Intrinsic 3D motion in covariant parton model}

The transverse momentum dependent parton distribution functions (TMDs) \cite%
{tmds,Mulders:1995dh} open the new way to a more complete understanding of
the quark-gluon structure of the nucleon. We studied this topic in our
recent papers \cite%
{Efremov:2009ze,Avakian:2009jt,Zavada:2009sk,Efremov:2009vb}. We have shown,
that requirements of symmetry (Lorentz invariance combined with rotationally
symmetric parton motion in the nucleon rest frame) applied in the covariant
parton model imply the relations between integrated unpolarized or polarized
distribution functions and their unintegrated counterparts. Further part is
devoted to the discussion on the quark orbital angular momentum and its
relation to the pretzelosity distribution function.

\section{Transversal motion}

Formulation of the model in terms of the light--cone formalism is suggested
in \cite{Efremov:2009ze} and allows to compute the chiral-even leading-twist
TMDs which are defined \cite{Mulders:1995dh}\ by means of the light--front
correlators $\phi (x,\mathbf{p}_{T})_{ij}$ as:%
\begin{equation}
\frac{1}{2}\;\mathrm{tr}\left[ \gamma ^{+}\;\phi (x,\mathbf{p}_{T})\right]
=f_{1}(x,\mathbf{p}_{T})-\frac{\varepsilon ^{jk}p_{T}^{j}S_{T}^{k}}{M}%
\,f_{1T}^{\perp }(x,\mathbf{p}_{T}),  \label{e1}
\end{equation}%
\begin{equation}
\frac{1}{2}\mathrm{tr}\left[ \gamma ^{+}\gamma _{5}\phi (x,\mathbf{p}_{T})%
\right] =S_{L}g_{1}(x,\mathbf{p}_{T})+\frac{\mathbf{p}_{T}\mathbf{S}_{T}}{M}%
g_{1T}^{\bot }(x,\mathbf{p}_{T}).  \label{e2}
\end{equation}%
In this section we assume mass of quark $m\rightarrow 0$. This assumption
substantially simplifies calculation within the model and seems to be in a
good agreement with experimental data -- in all model relations and rules,
where such comparison can be done. But in principle, more complicated
calculation with $m>0$ is possible \cite{Zavada:2002uz}.

The symmetry constraints applied in the model imply \cite%
{Zavada:2009sk,Efremov:2009vb} the relations between unintegrated
distribution and its integrated counterparts:%
\begin{equation}
f_{1}^{q}(x,\mathbf{p}_{T})=-\frac{1}{\pi M^{2}}\left( \frac{f_{1}^{q}(\xi )%
}{\xi }\right) ^{\prime },  \label{q11}
\end{equation}%
\begin{equation}
g_{1}^{q}(x,\mathbf{p}_{T})=\frac{2x-\xi }{\pi M^{2}\xi ^{3}}\left(
3g_{1}^{q}(\xi )+2\int_{\xi }^{1}\frac{g_{1}^{q}(y)}{y}dy-\xi \frac{d}{d\xi }%
g_{1}^{q}(\xi )\right) ,  \label{e3}
\end{equation}%
\begin{equation}
g_{1T}^{\bot q}(x,\mathbf{p}_{T})=\frac{2}{\pi M^{2}\xi ^{3}}\left(
3g_{1}^{q}(\xi )+2\int_{\xi }^{1}\frac{g_{1}^{q}(y)}{y}dy-\xi \frac{d}{d\xi }%
g_{1}^{q}(\xi )\right) ,  \label{e4}
\end{equation}%
where%
\begin{equation}
\xi =x\left( 1+\left( \frac{\mathbf{p}_{T}}{Mx}\right) ^{2}\right) .
\label{e5}
\end{equation}%
The time-reversal odd Sivers distribution function\ $f_{1T}^{\perp }$
requires explicit gluon degrees of freedom and is absent in our approach.
Apparently, the last two functions are related:%
\begin{equation}
\frac{g_{1}^{q}(x,\mathbf{p}_{T})}{g_{1T}^{\bot q}(x,\mathbf{p}_{T})}=\frac{x%
}{2}\left( 1-\left( \frac{\mathbf{p}_{T}}{Mx}\right) ^{2}\right) .
\label{e6}
\end{equation}%
Notice that from this relation the "Wandzura-Wilczek-type approximation" 
\cite{Avakian:2007mv} follows:%
\begin{equation}
g_{1T}^{\perp (1)q}(x)=x\int_{x}^{1}\frac{g_{1}^{q}(y)}{y}dy.  \label{e6a}
\end{equation}
Now, using the input distributions $f_{1}^{q}(x)$ and $g_{1}^{q}(x)$ one can
calculate corresponding TMDs.

\subsection{\textbf{Unpolarized distribution functions}}

\begin{figure}[tbp]
\includegraphics[width=12cm]{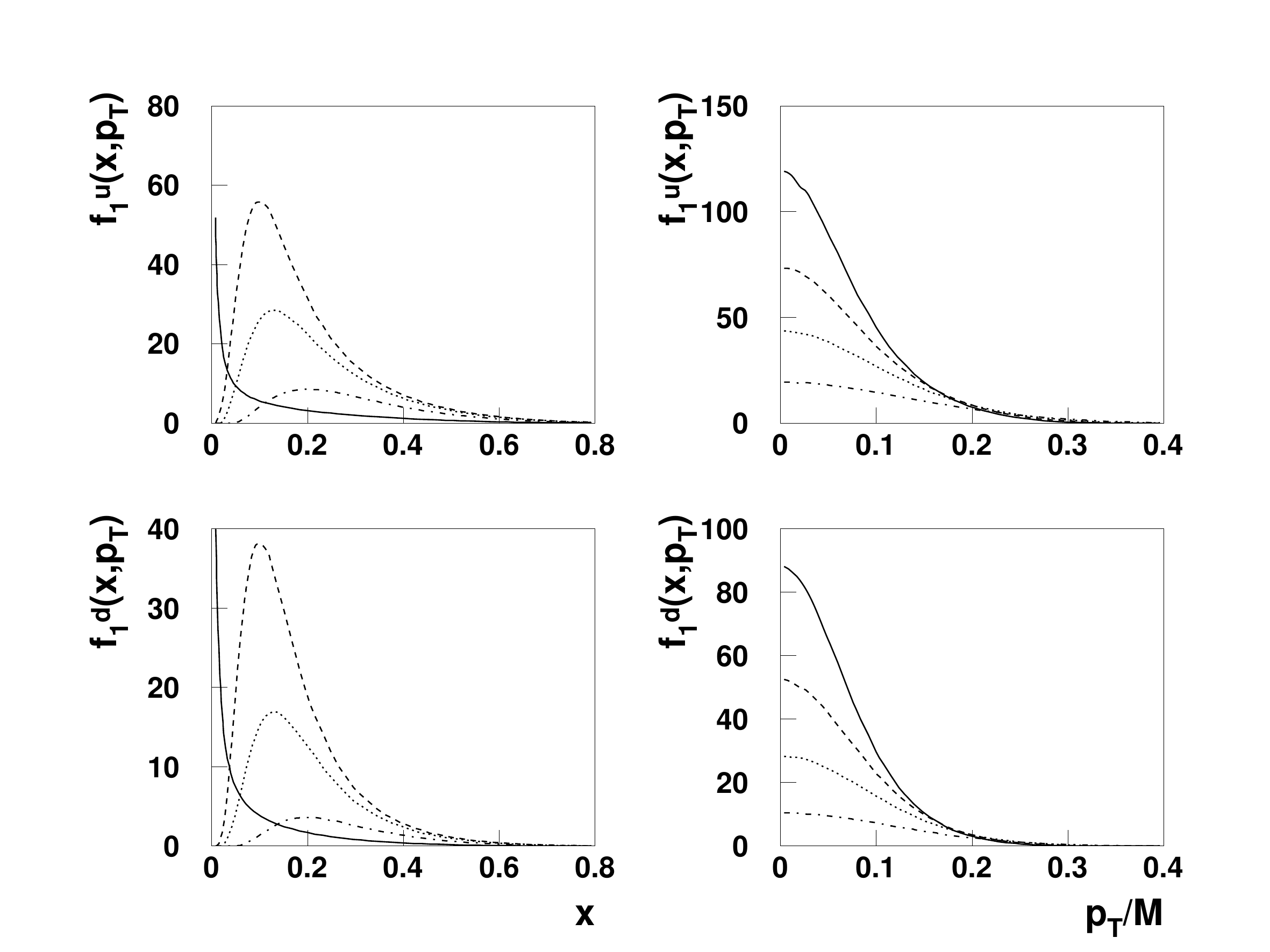}
\caption{Transverse momentum dependent unpolarized distribution functions
for $u$ (\textit{upper figures}) and $d-$quarks (\textit{lower figures}). 
\textbf{Left part}: dependence on $x$ for $p_{T}/M=0.10,0.13,0.20$ is
indicated by dash, dotted and dash-dot curves; solid curve correspods to the
integrated distribution $f_{1}^{q}(x)$. \textbf{Right part}: dependence on $%
p_{T}/M$ $\ $for $x=0.15,0.18,0.22,0.30$ is indicated by solid, dash, dotted
and dash-dot curves.}
\label{ff1}
\end{figure}

\begin{figure}[tbp]
\includegraphics[width=12cm]{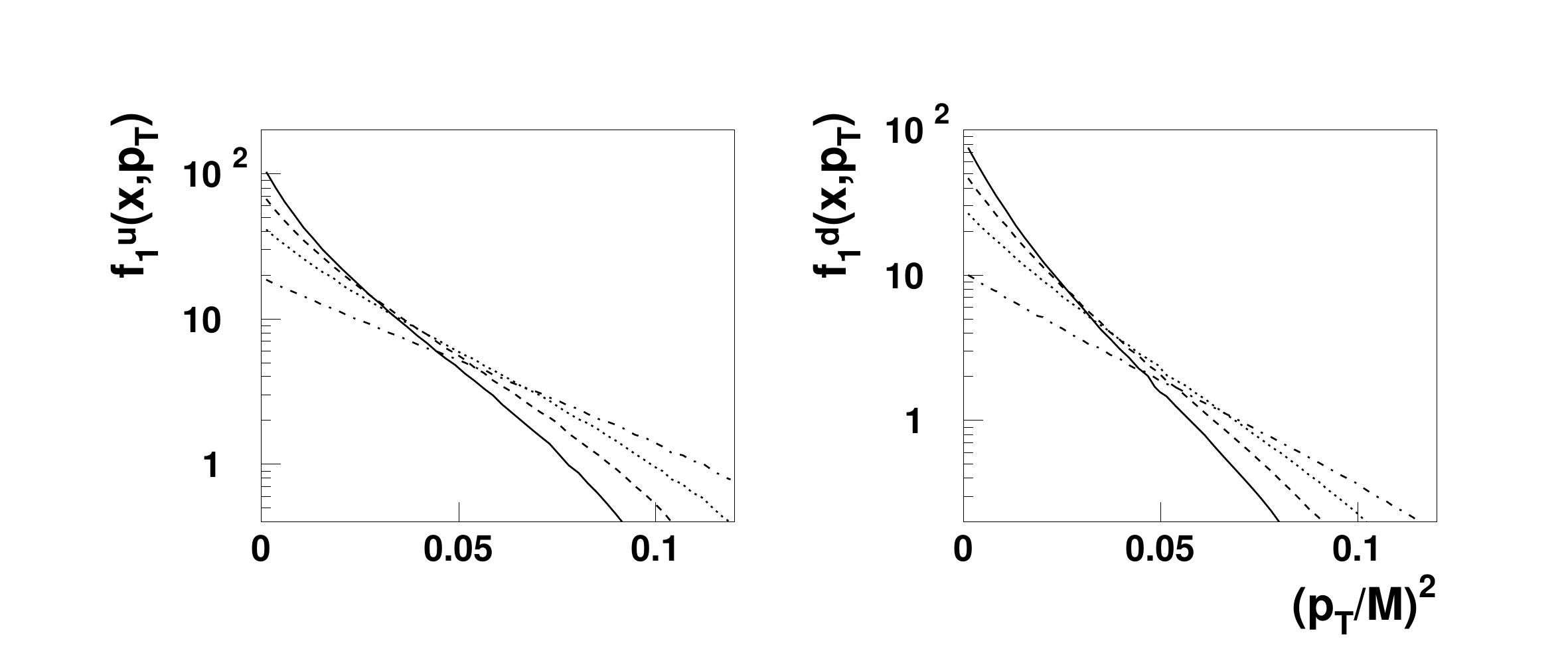}
\caption{Transverse momentum dependent unpolarized distribution functions
for $u$ and $d-$quarks. Dependence on $\left( p_{T}/M\right) ^{2}$ $\ $for $%
x=0.15,0.18,0.22,0.30$ is indicated by solid, dash, dotted and dash-dot
curves.}
\label{ff2}
\end{figure}

For the unpolarized input we used the standard PDF parameterization \cite%
{Martin:2004dh} (LO at the scale $4GeV^{2}$).\ In Fig. \ref{ff1} we have
results obtained from relation (\ref{q11}) for $u$ and $d-$quarks. 
The right part of this figure is shown again, but in different scale in Fig %
\ref{ff2}. 
One can observe the following:

\textit{i)} For fixed $x$ the  $p_{T}-$ distributions are very close to the
Gauss Ansatz $f_{1}^{q}(x,p_{T})\propto \exp \left( -p_{T}^{2}/\left\langle
p_{T}^{2}\right\rangle \right) .$ This is interesting result, since the
Gaussian shape is supported by phenomenology \cite{Schweitzer:2010tt}.

\textit{ii)} The width $\left\langle p_{T}^{2}\right\rangle $ depends on $x$%
. This result reflects to the fact, that in our approach, due to rotational
symmetry, the parameters $x$ and $p_{T}$ are not independent.

\textit{iii)} Figures suggest the typical values of transversal momenta, $%
\left\langle p_{T}^{2}\right\rangle \approx 0.01GeV^{2}$ or $\left\langle
p_{T}\right\rangle \approx 0.1GeV$. These values correspond to the estimates
based on the different analyses of the structure function $F_{2}(x,Q^{2})$ 
\cite{Zavada:2009sk}. On the other hand, much larger values $\langle
p_{T}^{2}\rangle \sim 0.4GeV^{2}$ are inferred from SIDIS data referring to
comparable scales \cite{Schweitzer:2010tt}, see also \cite%
{Anselmino:2005nn,Collins:2005ie}. Note also that in the statistical model
of TMDs \cite{Bourrely:2005tp} the parameter $\left\langle
p_{T}\right\rangle $ may be interpreted as an effective \cite%
{Cleymans:2010aa} temperature of partonic "ensemble". In turn, it may be
compared to the lattice calculations \cite{Karsch:2001vs} of the QCD phase
transition temperature $T\approx 175$ MeV.

\subsection{\textbf{Polarized distribution functions}}

\begin{figure}[tbp]
\includegraphics[width=12cm]{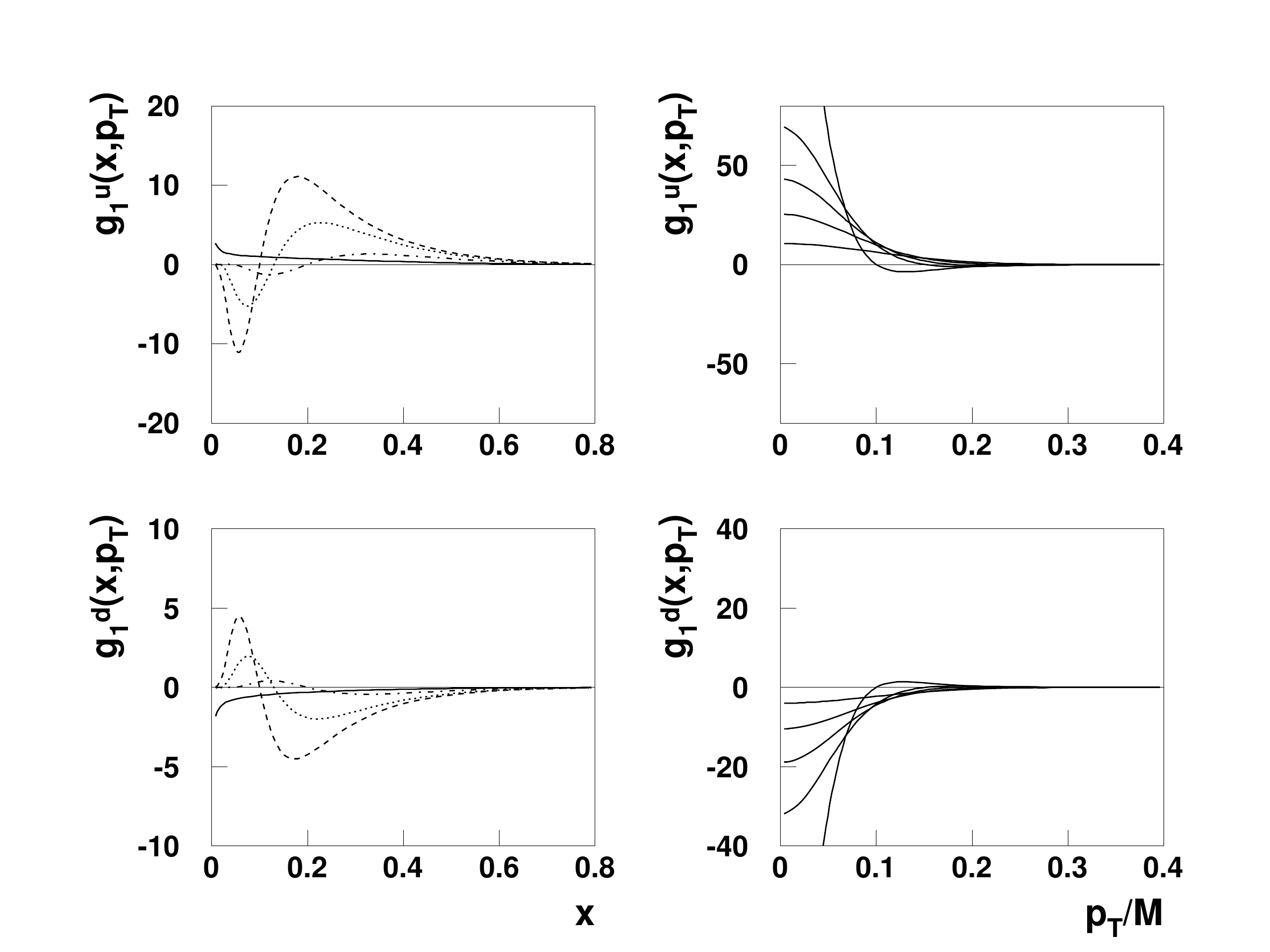}
\caption{Transverse momentum dependent polarized distribution functions for $%
u$ (\textit{upper figures}) and $d-$quarks (\textit{lower figures}). \textbf{%
Left part}: dependence on $x$ for $p_{T}/M=0.10,0.13,0.20$ is indicated by
dash, dotted and dash-dot curves; solid curve correspods to the integrated
distribution $g_{1}^{q}(x)$. \textbf{Right part}: dependence on $p_{T}/M$ $\ 
$for $x=0.10,0.15,0.18,0.22,0.30$ from top to down for $u-$quarks, and the
same symmetrically for $d-$quarks.}
\label{ff3}
\end{figure}

With the use of standard input \cite{lss} on \ $g_{1}^{q}(x)=\Delta q(x)/2$
to the relation (\ref{e3}) we obtain the curves $g_{1}^{q}(x,p_{T})$
displayed in Fig. \ref{ff3}. 
Let us remark, that the curves change the sign at the point $p_{T}=Mx$. This
change is due to the term%
\begin{equation}
2x-\xi =x\left( 1-\left( \frac{p_{T}}{Mx}\right) ^{2}\right) =2\tilde{p}%
_{1}/M  \label{m23}
\end{equation}%
in relation (\ref{e3}). This term is proportional to the quark longitudinal
momentum $\tilde{p}_{1}$ in the proton rest frame, which is defined by given 
$x$\ and $p_{T}$, see \cite{Zavada:2009sk}. It means, that sign of the $%
g_{1}^{q}(x,p_{T})$ is controlled by sign of the $\tilde{p}_{1}$. In fact,
there is some similarity to the function $g_{2}^{q}(x)$, which also changes
sign. The covariant parton model implies relation, which in the nucleon rest
frame read \cite{Zavada:2007ww}:%
\begin{equation}
g_{2}(x)=-\frac{1}{2}\int \Delta G(p_{0})\left( p_{1}+\frac{%
p_{1}^{2}-p_{T}^{2}/2}{p_{0}+m}\right) \delta \left( \frac{p_{0}+p_{1}}{M}%
-x\right) \frac{d^{3}p}{p_{0}}.  \label{sp11}
\end{equation}%
The $\delta -$function means, that large $x$ is correlated with great and
positive $p_{1}$ and on contrary the low $x$ with great but negative $p_{1}$%
. The kinematic term inside the integral changes the sign between the
extreme values of $\ p_{1}$, that is why the $g_{2}(x)$\ changes the sign.
Let us remark, that estimate of the $g_{2}(x)$ based on the relation (\ref%
{sp11}) well agrees \cite{Zavada:2002uz} with the experimental data.

\section{Orbital motion}

In the framework of covariant parton model we demonstrated that the 3D
picture of parton momenta inside the nucleon is a necessary input for
consistent accounting for quark OAM \cite{Zavada:2007ww}. Let us repeat the
main arguments. According to the rules of quantum mechanics the total
angular momentum  (in our case of a single quark) consists of the orbital
and spin part $\mathbf{j=l+s}$ and in relativistic case the $\mathbf{l}$ and 
$\mathbf{s}$\ are not conserved separately, but only the total angular
momentum $\mathbf{j}$ is conserved. General solution of Dirac equation for $%
j=j_{z}=1/2$ reads:%
\begin{equation}
\Psi \left( \mathbf{p}\right) =\int a_{k}\psi _{kjlj_{z}}\left( \mathbf{p}%
\right) dk;\quad \int a_{k}^{\star }a_{k}dk=1,  \label{sp21}
\end{equation}%
where%
\begin{equation}
\psi _{kjlj_{z}}\left( \mathbf{p}\right) =\frac{\delta (p-k)}{p\sqrt{8\pi
p_{0}}}\left( 
\begin{array}{c}
\sqrt{p_{0}+m}\left( 
\begin{array}{c}
1 \\ 
0%
\end{array}%
\right)  \\ 
-\sqrt{p_{0}-m}\left( 
\begin{array}{c}
\cos \theta  \\ 
\sin \theta \exp \left( i\varphi \right) 
\end{array}%
\right) 
\end{array}%
\right) .  \label{sp22}
\end{equation}%
The average spin contribution to the total angular momentum is defined as%
\begin{equation}
\left\langle s_{z}\right\rangle =\int \Psi ^{\dagger }\left( \mathbf{p}%
\right) \Sigma _{z}\Psi \left( \mathbf{p}\right) d^{3}p;\qquad \Sigma _{z}=%
\frac{1}{2}\left( 
\begin{array}{cc}
\sigma _{z} & \cdot  \\ 
\cdot  & \sigma _{z}%
\end{array}%
\right) ,  \label{sp26}
\end{equation}%
which implies

\begin{equation}
\left\langle s_{z}\right\rangle =\int a_{p}^{\star }a_{p}\frac{\left(
p_{0}+m\right) +\left( p_{0}-m\right) \left( \cos ^{2}\theta -\sin
^{2}\theta \right) }{16\pi p^{2}p_{0}}d^{3}p=\frac{1}{2}\int a_{p}^{\star
}a_{p}\left( \frac{1}{3}+\frac{2m}{3p_{0}}\right) dp.  \label{sp24}
\end{equation}%
Since\ $\left\langle s_{z}\right\rangle +\left\langle l_{z}\right\rangle
=j_{z}=1/2$, this relation implies for the orbital momentum:%
\begin{equation}
\left\langle l_{z}\right\rangle =\frac{1}{3}\int a_{p}^{\star }a_{p}\left( 1-%
\frac{m}{p_{0}}\right) dp.  \label{sp25}
\end{equation}%
In relativistic case, when $m\ll p_{0}$ in the nucleon rest frame, the role
of OAM for generating nucleon spin is dominant,%
\begin{equation}
\left\langle s_{z}\right\rangle \rightarrow 1/6,\qquad \left\langle
l_{z}\right\rangle \rightarrow 1/3.  \label{e15}
\end{equation}%
This result is related to the state $j=j_{z}=1/2$, where the axis $z$
represents direction of polarization. If the same state is polarized in any
other direction, then $\ -1/2<\left\langle j_{z}\right\rangle <1/2,$ but
still it holds%
\begin{equation}
\left\langle j_{z}\right\rangle =\left\langle s_{z}\right\rangle
+\left\langle l_{z}\right\rangle ,\qquad \left\langle l_{z}\right\rangle
=2\left\langle s_{z}\right\rangle .  \label{e16}
\end{equation}%
In the covariant parton model we identify $\left\langle s_{z}\right\rangle $
and $\left\langle l_{z}\right\rangle $ with the quark spin and orbital
momentum, so the sum over all quarks%
\begin{equation}
J_{z}^{quark}=\sum_{q}\left\langle j_{z}^{q}\right\rangle   \label{e17}
\end{equation}%
gives the total quark contribution to the nucleon spin. Due to (\ref{e16}),
only $1/3$ of this sum is generated by quark spins.

Now let us consider another representation of the quark spins and orbital
momenta. The spin contribution of quarks inside the nucleon to its spin is
defined as%
\begin{equation}
\left\langle s^{q}\right\rangle =\int g_{1}^{q}\left( x\right) dx.
\label{e7}
\end{equation}%
It has been suggested recently \cite{Avakian:2010br,She:2009jq}, that the
pretzelosity distribution $h_{1T}^{\perp (1)q}\left( x\right) $ is related
to the quark orbital momentum as%
\begin{equation}
\left\langle l^{q}\right\rangle =-\int h_{1T}^{\perp (1)q}\left( x\right) dx.
\label{e8}
\end{equation}%
On the other hand, as we showed in \cite{Efremov:2009ze} (Eq.22), expression
for pretzelosity in the covariant model reads%
\begin{equation}
h_{1T}^{\perp q}\left( x,p_{T}\right) =-M^{2}\int \frac{\Delta G(p_{0})}{%
p_{0}+m}\delta \left( \frac{p_{0}+p_{1}}{M}-x\right) \frac{dp_{1}}{p_{0}},
\label{e9}
\end{equation}%
from which we obtain the (1) -- moment%
\begin{equation}
h_{1T}^{\perp (1)q}\left( x\right) =\int \frac{p_{T}^{2}}{2M^{2}}%
h_{1T}^{\perp q}\left( x,p_{T}\right) d^{2}p_{T}=-\frac{1}{2}\int \Delta
G(p_{0})\frac{p_{T}^{2}}{p_{0}+m}\delta \left( \frac{p_{0}+p_{1}}{M}%
-x\right) \frac{d^{3}p}{p_{0}}  \label{e10}
\end{equation}%
and%
\begin{equation}
\int h_{1T}^{\perp (1)q}\left( x\right) dx=-\frac{1}{2}\int \Delta G(p_{0})%
\frac{p_{T}^{2}}{p_{0}+m}\frac{d^{3}p}{p_{0}}.  \label{e11}
\end{equation}%
After replacing $p_{T}^{2}\rightarrow \frac{2}{3}\left\vert \mathbf{p}%
\right\vert ^{2}$ and $\left\vert \mathbf{p}\right\vert ^{2}=p_{0}^{2}-m^{2%
\text{ }}$ one gets%
\begin{equation}
-\int h_{1T}^{\perp (1)q}\left( x\right) dx=\frac{1}{3}\int \Delta
G(p_{0})\left( 1-\frac{m}{p_{0}}\right) d^{3}p.  \label{e12}
\end{equation}%
For helicity the covariant model gives the relation%
\begin{equation}
g_{1}(x)=\frac{1}{2}\int \Delta G(p_{0})\left( m+p_{1}+\frac{p_{1}^{2}}{%
p_{0}+m}\right) \delta \left( \frac{p_{0}+p_{1}}{M}-x\right) \frac{d^{3}p}{%
p_{0}},  \label{e13}
\end{equation}%
which implies%
\begin{equation}
\int g_{1}^{q}\left( x\right) dx=\frac{1}{2}\int \Delta G_{q}\left(
p_{0}\right) \left( \frac{1}{3}+\frac{2m}{3p_{0}}\right) d^{3}p.  \label{e14}
\end{equation}%
We can arrange the two sets of results for average spin and orbital momentum
calculated by means:

1.\textit{\ wavefunctions and operators} (Eqs.(\ref{sp24}),(\ref{sp25})):

\[
\begin{tabular}{|c|c|}
\hline
$\left\langle s^{q}\right\rangle $ & $\left\langle l^{q}\right\rangle $ \\ 
\hline
\multicolumn{1}{|l|}{$\quad \frac{1}{2}\int a_{p}^{\ast }a_{p}\left( \frac{1%
}{3}+\frac{2m}{3p_{0}}\right) dp\quad $} & \multicolumn{1}{|l|}{$\quad \frac{%
1}{3}\int a_{p}^{\ast }a_{p}\left( 1-\frac{m}{p_{0}}\right) dp\quad $} \\ 
\hline
\end{tabular}%
\]

2. \textit{structure functions and probabilistic distributions} (Eqs.(\ref%
{e12}),(\ref{e14})):\ 
\[
\ 
\begin{tabular}{|c|c|}
\hline
$\int g_{1}^{q}\left( x\right) dx$ & $-\int h_{1T}^{\perp (1)q}\left(
x\right) dx$ \\ \hline
\multicolumn{1}{|l|}{$\frac{1}{2}\int \Delta G_{q}\left( p_{0}\right) \left( 
\frac{1}{3}+\frac{2m}{3p_{0}}\right) d^{3}p$} & \multicolumn{1}{|l|}{$\frac{1%
}{3}\int \Delta G_{q}\left( p_{0}\right) \left( 1-\frac{m}{p_{0}}\right)
d^{3}p$} \\ \hline
\end{tabular}%
\]%
Obviously, if we identify probabilities%
\begin{equation}
a_{p}^{\ast }a_{p}dp\Leftrightarrow \Delta G_{q}\left( p_{0}\right)
d^{3}p;\quad \Delta G_{q}\left( p_{0}\right) =G_{q}^{+}\left( p_{0}\right)
-G_{q}^{-}\left( p_{0}\right)   \label{e14z}
\end{equation}%
then the table implies, that relation (\ref{e8}) between orbital momentum
and pretzelosity is valid also in the covariant model.\medskip 

{\footnotesize \textbf{Acknowledgements.} A.~E. and O.~T. are supported by
the Grants RFBR 09-02-01149 and 07-02-91557, RF MSE RNP 2.1.1/2512(MIREA)
and (also P.Z.) Votruba-Blokhitsev Programs of JINR. P.~Z. is supported by
the project AV0Z10100502 of the Academy of Sciences of the Czech Republic.
The work was supported in part by DOE contract DE-AC05-06OR23177.}

\end{document}